\documentclass[conference]{IEEEtran}
\usepackage{float}
\IEEEoverridecommandlockouts
\usepackage{cite}
\usepackage{amsmath,amssymb,amsfonts}
\usepackage{algorithmic}
\usepackage{graphicx}
\usepackage{textcomp}
\usepackage{xcolor}
\usepackage{booktabs}
\usepackage{multirow}
\usepackage{fancyhdr}
\usepackage{url}
\def\BibTeX{{\rm B\kern-.05em{\sc i\kern-.025em b}\kern-.08em
    T\kern-.1667em\lower.7ex\hbox{E}\kern-.125emX}}
 
\usepackage{fancyhdr}
\fancypagestyle{firstpage}{
    \fancyfoot[L]{*979-8-3195-1706-7/26/\$31.00~\copyright2026 IEEE \hfill}
}
 
\begin{document}
 
\title{Optimization of a Hybrid Solar-Biogas-Battery Energy Storage Microgrid: A Case Study}
 
\author{
\IEEEauthorblockN{John Elomunait}
\IEEEauthorblockA{\textit{College of Engineering} \\
\textit{Carnegie Mellon University Africa}\\
Kigali, Rwanda  \\
jelomuna@andrew.cmu.edu}
 
\and
\IEEEauthorblockN{Jesse Thornburg}
\IEEEauthorblockA{
\textit{Department of Electrical and Computer Engineering}\\
Carnegie Mellon University, Pittsburgh, PA, USA \\
Kigali Collaborative Research Centre, Kigali, Rwanda \\
Scott Institute for Energy Innovation at CMU, PA, USA \\
jthornbu@andrew.cmu.edu}
}

\IEEEoverridecommandlockouts
\IEEEpubid{\makebox[\columnwidth]{*979-8-3195-1706-7/26/\$31.00~\copyright2026 IEEE \hfill}
\hspace{\columnsep}\makebox[\columnwidth]{ }}
 
\maketitle
\thispagestyle{firstpage}

\begin{abstract}
Low-income regions like eastern Uganda often lack reliable grid power. This paper presents the techno-economic optimization of the Soroti Integrated Renewable Energy Microgrid (SIREM), a hybrid renewable system for grid-access expansion where Soroti District household grid access stands at 16.7\%. The microgrid integrates concentrating solar power (CSP) with thermal storage, biogas combined cycle generation, and lithium iron phosphate battery storage. Using Xendee optimization software with six years of NASA POWER data, the optimal configuration comprises a 37 MW CSP field, 70 MWh molten-salt storage, a 5 MW CSP turbine, a 6 MW biogas CCGT, and a 10 MWh battery system. The design delivers 61.3 GWh annually with a 42.91\% IRR, 2-year payback, and levelized cost of electricity of 0.1991 USD/kWh generation (0.3752 USD/kWh load-served), achieving 32.2\% total savings versus grid-only supply while exporting 28.8 GWh annually, demonstrating the viability of hybrid renewable microgrids for grid expansion in sub-Saharan Africa.
\end{abstract}


\section{Introduction}

Sub-Saharan Africa faces the world's most acute electricity access deficit, with approximately 600 million people lacking grid connection as of 2024 \cite{murungi2025}. In Uganda, following the commissioning of the 600 MW Karuma Hydropower Plant, installed generation capacity reached approximately 2,098.2 MW by mid-2025. However, the 2024 census found that only 25.3\% of households rely on the national grid for lighting, with rural grid access at just 9.1\% \cite{murungi2025}. When including off-grid solar solutions, total household electricity access rises to 53.4\% \cite{UBOS2024}. Peak consumption of 985 MW is constrained not by generation but by an undersized distribution network, which also suffers from system losses of around 17.1\% \cite{trade.gov2025}.

For the combined area of Soroti District and Soroti City District, the 2024 census shows a weighted grid access of 16.7\% of households. \cite{ubos_soroti_district_2024}. The district hosts Soroti City (population 134,199), Soroti Regional Referral Hospital, Soroti University of Technology, and the government-designated Soroti Industrial and Business Park. The agricultural economy, with 102,267 cattle as of 2021, and substantial crop production, provides biogas feedstock while creating agro-processing electrification demand \cite{UBOS_livestock2021}.

Uganda’s Electricity Act (Cap. 157), administered by the Electricity Regulatory Authority (ERA), permits independent power producers to obtain generation and distribution licenses. ERA’s Renewable Energy Feed-in Tariff Phase 5 guidelines establish fixed tariffs for hydro and bagasse generation, as well as tariff ceilings for biogas, solar, and other renewable technologies \cite{ERA2021}. Current retail tariffs (Q1 2026) range from 0.069 USD/kWh (domestic lifeline) to 0.209 USD/kWh (domestic standard), with commercial at 0.151 USD/kWh \cite{ERA2026}.

This paper contributes to the wider microgrid-modeling literature by answering four research questions: (1) How to optimally size a hybrid CSP-biogas-battery configuration for a specific location? (2) What are the techno-economic performance metrics? (3) How does such a system compare to grid-only supply? (4) What is the environmental impact and alignment with national climate targets?

\section{System Design and Resource Assessment}

\subsection{Load Profile}

The SIREM system serves approximately 5,226 first-connection customers across residential, agricultural processing, institutional (hospital, university), commercial, and industrial users. These categories have markedly different demand profiles and peaks, so they are modeled differently \cite{ustun,thornburg}.

Total annual electricity demand is estimated at 40.8 GWh based on conservative per-customer consumption assumptions aligned with Uganda's consumption patterns for newly-connected households and small commercial users. Peak aggregate demand is modeled at approximately 7 MW during evening hours (18:00-22:00 local time).

\IEEEpubidadjcol

\subsection{Solar Resource Assessment}

\begin{figure*}
\centering
\includegraphics[width=0.73\textwidth]{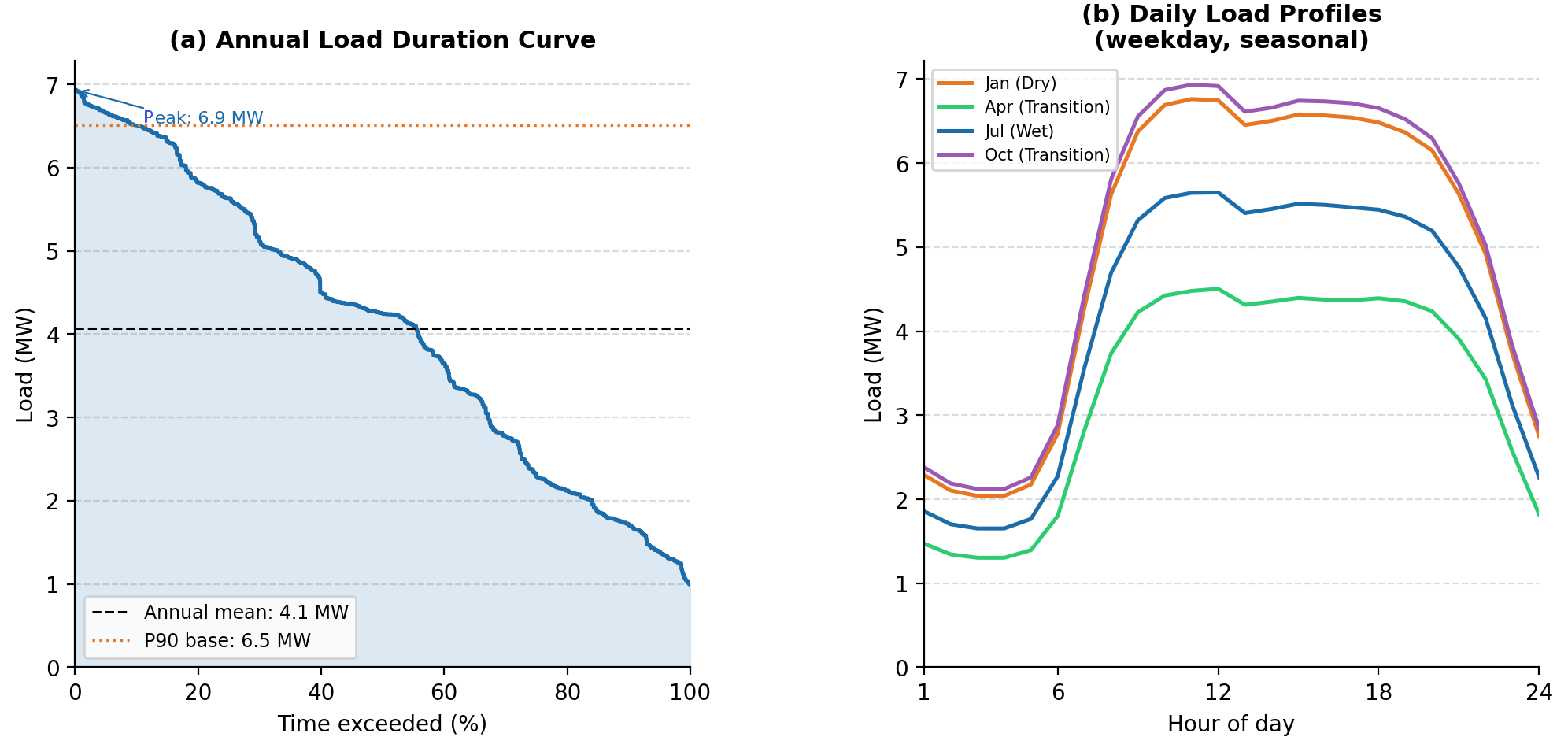}

\caption{(a) Annual load duration curve and (b) representative weekday load profiles for 
the SIREM site. The load duration curve (a) is derived from the electricity-only load data weighted by monthly day-counts; peak demand reaches 6.93~MW with an annual mean of 4.07~MW. The seasonal weekday profile (b) shows a pronounced morning ramp (07:00--10:00) and elevated afternoon plateau (12:00--17:00) due to commercial and industrial end-use patterns.}
\label{fig:load}
\end{figure*}

Six years of hourly NASA POWER data (January 2020 through January 2026)
for coordinates 1.627$^{\circ}$N, 33.526$^{\circ}$E yield a mean
direct normal irradiance of 4.60~kWh/m\textsuperscript{2}/day and
global horizontal irradiance of 5.89~kWh/m\textsuperscript{2}/day from
52,608 hourly records~\cite{NASA_POWER2026}. The site exhibits
relatively consistent year-round insolation with modest seasonal
variation, making it well-suited for CSP deployment. Direct normal
irradiance exceeds 4.0~kWh/m\textsuperscript{2}/day for over 300 days
each year.

\subsection{Biogas Resource Assessment}

Conservative biogas mass balance calculations based on regional cattle
population (102,267 head) and agricultural residues from rice, cassava,
and sorghum production estimate 30,000~m\textsuperscript{3}/day of
upgraded biomethane at 45\% collection efficiency. This planning figure
assumes cattle manure at 20~kg/animal/day (8\% total solids), methane
yield of 0.25~m\textsuperscript{3}~CH\textsubscript{4}/kg volatile
solids, and 95\% upgrading efficiency to pipeline-quality biomethane.

\begin{figure}[h]
\centering
\includegraphics[width=0.5\textwidth]{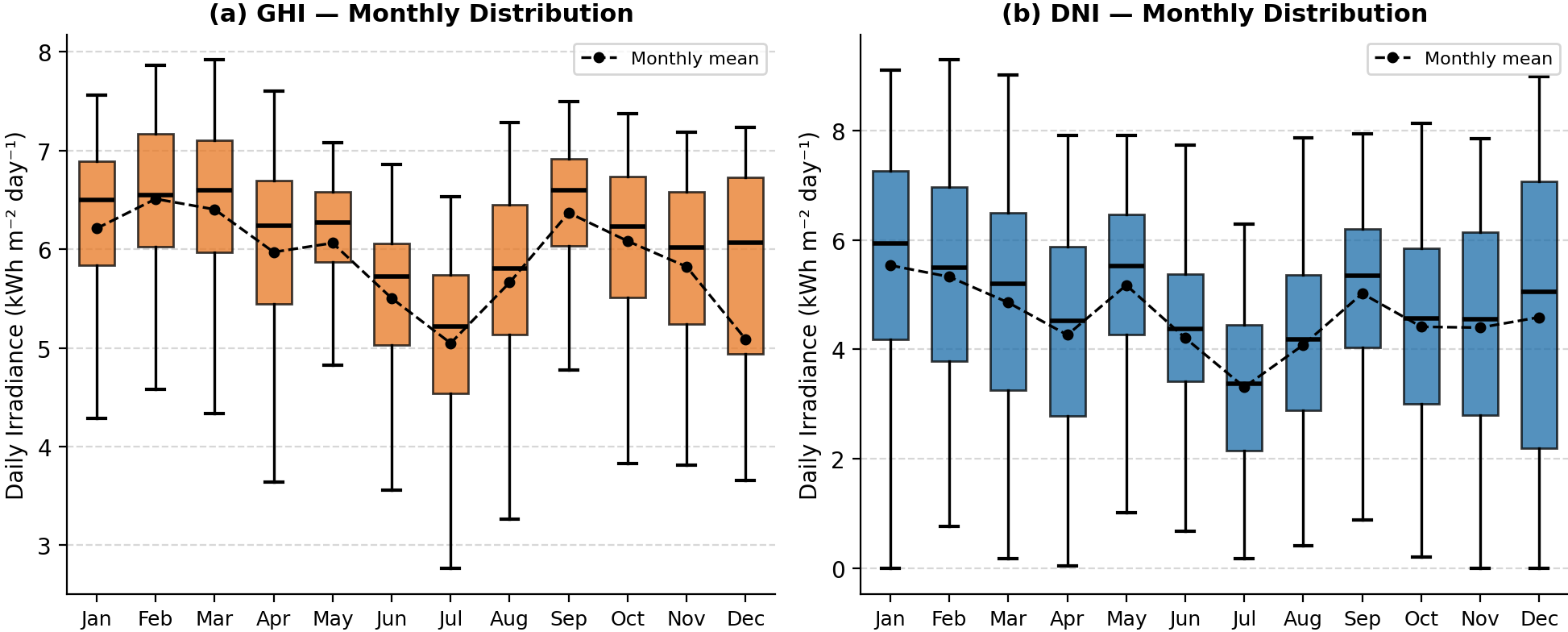}
\caption{Monthly direct normal irradiance and global horizontal irradiance 
distributions at Soroti site (2020--2026 NASA POWER data). 
Boxes show interquartile range of daily totals; whiskers extend to 
1.5$\times$IQR; dashed line connects monthly means.}
\label{fig:solar}
\end{figure}

This biogas production capacity is sufficient for 3.0 MW average combined cycle gas turbine dispatch with a 65\% fuel reserve margin, providing dispatchable generation for load-following and grid outage conditions.

\subsection{Technology Selection}

The CSP subsystem employs parabolic trough technology with molten-salt
thermal energy storage, selected for proven deployment in similar
tropical sites and technical maturity~\cite{Price2002,IRENA2013}.
Thermal storage enables dispatch decoupling from instantaneous solar
availability, providing generation into evening peak hours. The biogas
subsystem uses reciprocating engine CCGT technology optimized for
biogas fuel composition, providing fast-ramping dispatchable generation
and black-start capability. Lithium iron phosphate (LFP) battery
technology was selected for superior cycle life, thermal stability in
tropical conditions, and lower fire risk~\cite{thornburg26}, providing
sub-second frequency regulation and CSP output smoothing.

\section{Optimization Methodology}

\subsection{Mathematical Problem Formulation}

The Xendee optimization engine uses mixed-integer linear programming
(MILP) to simultaneously determine component sizing and operational
dispatch over a 25-year planning horizon with each year discretized into 8,760 hourly intervals. For MILP and economic analyses combined, Xendee is better than distribution and transmission-focused simulators like GridLab-D, PowerFactory, PLASMiS, or Adaptricity \cite{thesis2018, powerafrica2021, 9844342}.

\subsubsection{Decision Variables}

\textit{Continuous sizing variables}: CSP solar field capacity
$\bar{P}_{\text{CSP}}$~(MW$_{\text{th}}$), TES capacity
$\bar{E}_{\text{TES}}$~(MWh$_{\text{th}}$), CSP turbine rating
$\bar{P}_{\text{turb}}$~(MW$_{\text{e}}$), biogas CCGT capacity
$\bar{P}_{\text{BG}}$~(MW$_{\text{e}}$), and battery capacity
$\bar{E}_{\text{bat}}$~(MWh).

\textit{Continuous operational variables}: At each hour~$t$, unit
dispatch $p_k(t)$~(MW), TES state-of-charge
$E_{\text{TES}}(t)$~(MWh$_{\text{th}}$), battery SOC
$E_{\text{bat}}(t)$~(MWh), grid import $p_{\text{imp}}(t)$ and
export $p_{\text{exp}}(t)$~(MW), and non-served load
$p_{\text{NS}}(t)$~(MW).

\textit{Binary commitment variables}: $u_k(t)\in\{0,1\}$ indicating
on/off status of each dispatchable unit~$k$ at hour~$t$. Minimum
run-time constraints are not imposed given the 1-hour dispatch
resolution and the flexibility of reciprocating engine CCGT technology;
the binary variables enforce that dispatch is zero when a unit is
committed off, preventing fractional commitment artefacts in the
LP relaxation.

\subsubsection{Objective Function}

The MILP minimizes net present cost (NPC) over planning horizon
$\mathcal{Y}=\{1,\ldots,25\}$:
\begin{equation}
\min\;\text{NPC} = \text{CAPEX}_{\text{net}} +
\sum_{y\in\mathcal{Y}} \frac{\text{OPEX}_y - \text{REV}_y +
\text{REPL}_y}{(1+r)^y}
\label{eq:obj}
\end{equation}
where $r=0.11$ is the weighted average cost of capital (WACC) and
$\text{CAPEX}_{\text{net}}=\sum_k c_k^{\text{inv}}\bar{P}_k
-\tau_{\text{ITC}}\cdot\text{CAPEX}_{\text{elig}}$. The annual terms
are:
\begin{align}
\text{OPEX}_y &= \sum_{t}\!\left(c_{\text{fuel}}\,p_{\text{BG}}(t)
  + {\textstyle\sum_k} c_{\text{O\&M},k}\,p_k(t)
  + c_{\text{NS}}\,p_{\text{NS}}(t)\right)\!\Delta t
\label{eq:opex}\\[2pt]
\text{REV}_y  &= \sum_{t}\!\left(\pi_{\text{PPA}}\,
  p_{\text{on-site}}(t) + \pi_{\text{exp}}\,
  p_{\text{exp}}(t)\right)\!\Delta t
\label{eq:rev}
\end{align}
Here $\tau_{\text{ITC}}$ is the renewable energy investment tax credit
(30\% for battery and solar-side assets, 10\% for the biogas CCGT,
applied in accordance with Uganda ERA incentive
guidelines~\cite{ERA2021}), $c_{\text{fuel}}=0.015$~USD/kWh$_{\text{fuel}}$,
$\pi_{\text{PPA}}\in[0.12,0.18]$~USD/kWh, and
$\pi_{\text{exp}}=0.08$~USD/kWh (ERA REFIT Phase~5
rate~\cite{ERA2021}).

\subsubsection{Operational Constraints}

\noindent\textit{Nodal energy balance} ($\forall\,t$):
\begin{multline}
p_{\text{turb}}(t) + p_{\text{BG}}(t) + p_{\text{dis}}(t)
  + p_{\text{imp}}(t) = \\
P_{\text{load}}(t) + p_{\text{chg}}(t) + p_{\text{exp}}(t)
  + p_{\text{NS}}(t)
\label{eq:balance}
\end{multline}

\noindent\textit{TES dynamics} ($\sigma$~= thermal loss rate,
$\eta_{\text{abs}}$~= solar field absorber efficiency,
$\eta_{\text{turb}}=0.15$ = gross thermal-to-electric conversion
efficiency of the Rankine cycle):
\begin{align}
E_{\text{TES}}(t) &= (1{-}\sigma)\,E_{\text{TES}}(t{-}1)
  + \eta_{\text{abs}}\,P_{\text{CSP}}(t)\,\Delta t
  - \tfrac{p_{\text{turb}}(t)}{\eta_{\text{turb}}}\,\Delta t
\label{eq:tes}\\[2pt]
0 &\leq E_{\text{TES}}(t) \leq \bar{E}_{\text{TES}}
\label{eq:tes_lim}
\end{align}

\noindent\textit{Battery dynamics}
($\eta_c$,~$\eta_d$~= charge/discharge efficiencies,
$\eta_{\text{rt}}=\eta_c\cdot\eta_d=0.90$~= round-trip efficiency):
\begin{align}
E_{\text{bat}}(t) &= E_{\text{bat}}(t{-}1)
  + \eta_c\,p_{\text{chg}}(t)\,\Delta t
  - \tfrac{p_{\text{dis}}(t)}{\eta_d}\,\Delta t
\label{eq:bat}\\[2pt]
0.20\,\bar{E}_{\text{bat}} &\leq E_{\text{bat}}(t)
  \leq 0.90\,\bar{E}_{\text{bat}}
\label{eq:bat_soc}
\end{align}

\noindent\textit{Capacity and commitment limits} ($\forall k,t$):
\begin{equation}
0 \leq p_k(t) \leq u_k(t)\cdot\bar{P}_k, \quad u_k(t)\in\{0,1\}
\label{eq:gen_lim}
\end{equation}

\noindent\textit{Land area constraint} (Soroti site), where
$\eta_{\text{opt}}$~= optical efficiency of the parabolic-trough
collector field (dimensionless) and $G_b$~= design-point beam
irradiance (W/m\textsuperscript{2}) used for aperture sizing:
\begin{equation}
A_{\text{CSP}} = \bar{P}_{\text{CSP}}\,/\,
(\eta_{\text{opt}}\cdot G_b) \leq 75{,}000~\text{m}^2
\label{eq:land}
\end{equation}

The study uses Xendee Expert Mode for a MILP formulation using a branch-and-bound solver, producing globally optimal sizing and dispatch decisions with
731,336 equations, 23,267 binary variables, and 370,975 continuous
variables, converging in 35~seconds.

\subsection{Model Configuration and Constraints}

Xendee Expert Mode was configured with a 25-year planning horizon
(2026--2051), 11\% WACC, 30\% corporate tax, 70\% debt at 8\% loan
interest over 20~years, renewable energy ITC (30\% battery,
10\% CCGT)~\cite{ERA2021}, and 2\%/yr inflation. Technology unit costs:
CSP field 802~USD/kW$_{\text{th}}$, TES 138.6~USD/kWh$_{\text{th}}$,
CSP turbine 2,200~USD/kW$_{\text{e}}$, biogas CCGT 1,400~USD/kW,
battery 370~USD/kWh. Technical constraints: 75,000~m\textsuperscript{2}
CSP field limit, 10~annual battery cycles maximum, bidirectional grid
interconnection, 10--20~ms islanding capability, and 15\% minimum
reserve. Renewable dispatch is prioritized; CCGT provides load-following, a battery provides short-duration firming, and the surplus is exported to the grid.

\section{Optimization Results}

\subsection{Optimal System Configuration}

Table~\ref{tab:optimal_config} presents the optimal system
configuration.

\begin{table}[h]
\caption{Optimal SIREM System Configuration}
\begin{center}
\begin{tabular}{lcc}
\toprule
\textbf{Technology} & \textbf{Capacity} & \textbf{CAPEX (kUSD)} \\
\midrule
CSP Solar Field & 37~MW$_{\text{th}}$ & 29,679 \\
CSP Thermal Storage & 70~MWh$_{\text{th}}$ & 9,700 \\
CSP Turbine & 5~MW$_{\text{e}}$ & 11,000 \\
Biogas CCGT & 6~MW$_{\text{e}}$ (3 units) & 8,400 \\
Battery ESS & 10~MWh & 3,700 \\
\midrule
\textbf{Total} & & \textbf{62,479} \\
\bottomrule
\end{tabular}
\label{tab:optimal_config}
\end{center}
\end{table}

The CSP fills 69,358~m\textsuperscript{2} land area
(92.5\% of the 75,000~m\textsuperscript{2} constraint). The battery
inverter is 4,750~kW as calculated (5~MW specified), yielding a 0.95 inverter-to-capacity ratio.

\begin{figure}[h]
\centering
\includegraphics[width=0.5\textwidth]{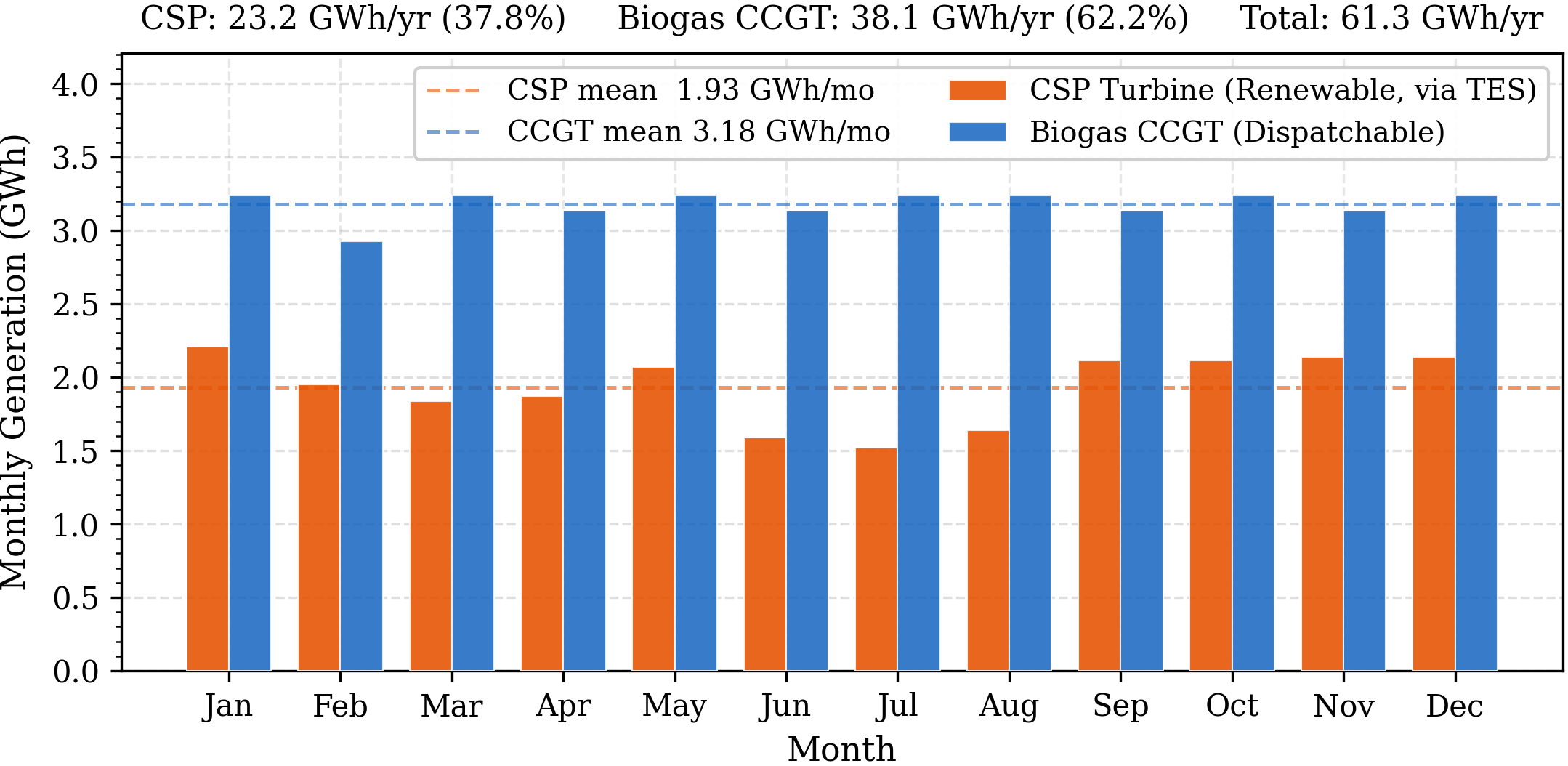}
\caption{Monthly on-site power generation: CSP field and biogas CCGT.}
\label{fig:generation}
\end{figure}

\subsection{Energy Balance and Performance Metrics}

The system generates 61.3~GWh/yr (38.1~GWh CCGT, 23.2~GWh CSP).
Of total site demand of 40.8~GWh, 32.5~GWh (79.6\%) is physically
served and 8.3~GWh (20.4\%) is shed as non-emergency load shedding,
accounting for the \$7.9M/yr shedding cost in Table~\ref{tab:cashflow}.
The 28.8~GWh exported represents surplus generation during grid-available
periods, closing the bus-bar energy balance
(32.5~+~28.8~+~0.014~losses~$\approx$~61.3~GWh).
CSP capacity factor is 19.67\%, CSP turbine 52.92\%, biogas CCGT
72.52\% (6,570~hr/yr). Battery completes 9~cycles/yr (87.6\% of limit).

\begin{figure*}
\centering
\includegraphics[width=\textwidth]{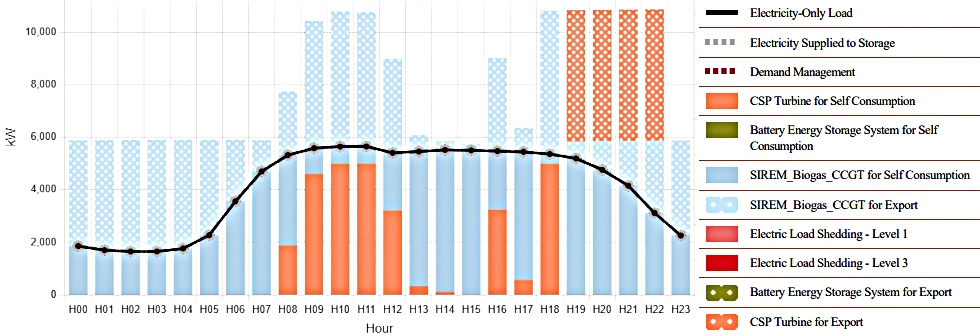}
\caption{Representative July daily electricity dispatch for the peak
demand day type. Power output (MW) by source.
CSP turbine dispatched hours H08--H14 from direct solar and H16--H22
from TES evening dispatch; biogas CCGT running continuously at
$\sim$4~MW with peaks when CSP is unavailable.}
\label{fig:dispatch}
\end{figure*}

\subsection{Financial Performance}

Table \ref{tab:financial} summarizes key financial performance indicators for the optimized SIREM system.

\begin{table}[h]
\caption{Financial Performance Metrics}
\begin{center}
\begin{tabular}{lc}
\toprule
\textbf{Metric} & \textbf{Value} \\
\midrule
Total CAPEX & 62.5 million USD \\
Equity (30\%) & 18.7 million USD \\
Debt (70\%) & 43.7 million USD \\
Generation-based LCOE & 0.1991 USD/kWh \\
Load-served LCOE & 0.3752 USD/kWh \\
OPEX savings & 69.7\% \\
Payback period & 2 years \\
Break-even year & 3 years \\
IRR (25-year) & 42.91\% \\
NPV (25-year) & 48.8 million USD \\
\midrule
Annual costs (reference) & 18.0 million USD \\
Annual costs (optimized) & 12.2 million USD \\
\textbf{Total savings (\%)} & \textbf{32.2\%} \\
\bottomrule
\end{tabular}
\label{tab:financial}
\end{center}
\vspace{-6mm}
\end{table}

The generation-based LCOE of 0.1991 USD/kWh compares favorably to Uganda's Q1 2026 domestic standard tariff of 0.209 USD/kWh and commercial tariff of 0.151 USD/kWh \cite{ERA2026}. The load-served LCOE of 0.3752 USD/kWh reflects the distribution cost of serving dispersed rural loads.

The system achieves positive cumulative cash flow by year 3 and reaches an aggregate return on investment of 336\% by year 25. The debt service coverage ratio maintains 3.0 throughout the debt amortization period (years 1-20), well above the 1.25 minimum typically required by project lenders.

\subsection{Cash Flow Analysis}

\begin{table}[t]
\caption{Cash Flow Projection: Savings Relative to Grid-Only
Reference (Thousand USD). Positive values are savings; negative
values are additional costs versus the reference.}
\begin{center}
\scriptsize
\setlength{\tabcolsep}{3.5pt}
\begin{tabular}{p{2.8cm} r r r r r r}
\toprule
\textbf{Item} & \textbf{Y0} & \textbf{Y1} & \textbf{Y2} & \textbf{Y3} & \textbf{Y4} \\
\midrule
Total OPEX savings & 0.0 & 12,551.0 & 12,551.0 & 12,551.0 & 12,551.0 \\
\midrule
CAPEX -- CSP Field & (8,903.7) & (2,116.0) & (2,116.0) & (2,116.0) & (2,116.0) \\
CAPEX -- TES & (2,910.0) & (691.6) & (691.6) & (691.6) & (691.6) \\
CAPEX -- Turbine & (3,300.0) & (784.3) & (784.3) & (784.3) & (784.3) \\
CAPEX -- Battery & (1,110.0) & (263.8) & (263.8) & (263.8) & (263.8)\\
CAPEX -- Biogas CCGT & (2,520.0) & (598.9) & (598.9) & (598.9) & (598.9)\\
\textbf{Total CAPEX difference} & \textbf{(18,743.7)} & \textbf{(4,454.5)} & \textbf{(4,454.5)} & \textbf{(4,454.5)} & \textbf{(4,454.5)} \\
\midrule
Renew. energy ITC & 0.0 & 1,672.5 & 0.0 & 0.0 & 0.0 \\
\midrule
Replacement -- TES & 0.0 & (385.0) & (385.0) & (385.0) & (385.0) \\
Replacement -- Battery & 0.0 & (175.0) & (175.0) & (175.0) & (175.0) \\
\textbf{Total replacement costs} & \textbf{0.0} & \textbf{(560.0)} & \textbf{(560.0)} & \textbf{(560.0)} & \textbf{(560.0)} \\
\midrule
\textbf{Net CF (non-discounted)} & \textbf{(18,743.7)} & \textbf{9,209.0} & \textbf{7,536.5} & \textbf{7,536.5} & \textbf{7,536.5} \\
\textbf{Net CF (discounted)} & \textbf{(18,743.7)} & \textbf{8,296.4} & \textbf{6,116.8} & \textbf{5,510.6} & \textbf{4,964.5} \\
\midrule
\textbf{Cumul. CF (non-disc.)} & \textbf{(18,743.7)} & \textbf{(9,534.7)} & \textbf{(1,998.2)} & \textbf{5,538.3} & \textbf{13,074.8} \\
\textbf{Cumul. CF (disc.)} & \textbf{(18,743.7)} & \textbf{(8,589.8)} & \textbf{(1,621.8)} & \textbf{4,049.6} & \textbf{8,612.8} \\
\bottomrule
\end{tabular}
\label{tab:cashflow}
\end{center}
\vspace{-6mm}
\end{table}

Table~\ref{tab:cashflow} shows savings relative to the grid-only
reference (positive = saving). Annual OPEX savings of 12.6~million USD
from Year~1 are driven by on-site generation displacing retail tariff
purchases. The upfront equity outlay (18.7~million USD) is partially
offset by the 1.7~million USD renewable energy ITC in Year~1~\cite{ERA2021}.
Cumulative cash flow turns positive in Year~3 on both a non-discounted
(5,538~kUSD) and discounted basis at 11\% WACC (4,050~kUSD), yielding
48.8~million USD NPV over 25~years.

The optimized system exports 28.8~GWh/year to the grid and achieves \$12.6M/yr OPEX savings (69.7\%), \$5.8M/yr total annualized savings (32.2\% including capital), and increased reliability (via black-start islanding) over a grid-only reference costing \$18.0M/year. The comparison assumes 100\% grid availability, which is conservative given documented unreliability of the grid in Eastern Uganda~\cite{ERA2023}.

\section{Environmental Impact Assessment}

Annual CO\textsubscript{2} emissions from CCGT biogas combustion total
13,102~metric~tons, while CSP grid exports displace 1,438~metric~tons
of grid-side generation (using Xendee's modeled grid emission factor
for Uganda). The net system CO\textsubscript{2} as reported by the
Xendee optimizer is therefore 11,665~metric~tons per year
(13,102~$-$~1,438~$=$~11,664, rounded). Against the reference
grid-only baseline of 11,400~metric~tons configured in the Xendee
model~\cite{Uganda_NDC2022}, the optimized system produces
265~metric~tons (2.3\%) more CO\textsubscript{2} on a direct
operational basis. The dominant climate
benefit of the project is therefore not direct CO\textsubscript{2}
reduction but rather the prevention of fugitive methane emissions from
unmanaged manure decomposition. Biogas capture prevents approximately
82,500~metric~tons CO\textsubscript{2}e per year from entering the
atmosphere (IPCC AR6 GWP$_{20}=82.5$~\cite{IPCC2021}), a benefit
approximately 312 times larger than the direct emission increase
versus the reference case.

\begin{figure}[h]
\centering
\includegraphics[width=0.5\textwidth]{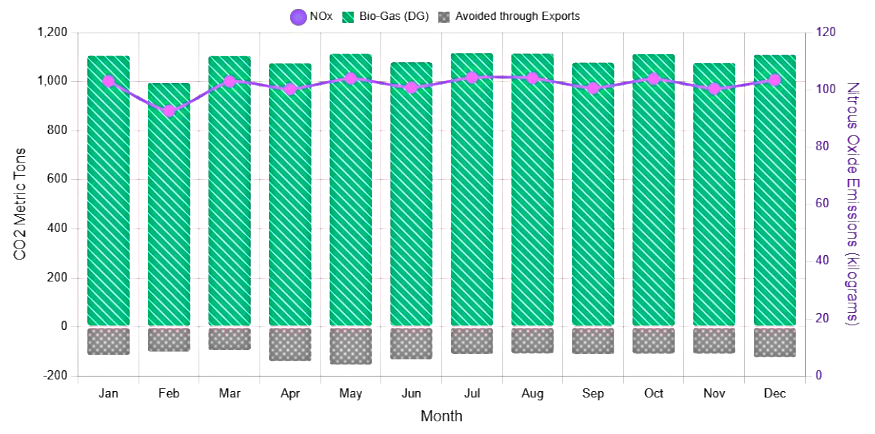}
\caption{Monthly CO\textsubscript{2} emissions from biogas combustion
(dark bars, 13,102~metric~tons/yr gross) and emissions avoided through
CSP grid exports (hatched bars, 1,438~metric~tons/yr as modeled by
Xendee). Net system CO\textsubscript{2} = 11,665~metric~tons/yr,
which is 265~metric~tons (2.3\%) above the 11,400~metric~ton
grid-only reference baseline.}
\label{fig:emissions}
\vspace{-6pt} 
\end{figure}

Uganda's NDC targets 22\% GHG reduction and 61\% renewable electricity
by 2030~\cite{Uganda_NDC2022}. SIREM achieves a 37.8\% renewable
fraction on-site (47.6\% including exports). Its principal climate
benefit is avoiding approximately 82,500~metric~tons
CO\textsubscript{2}e/yr through methane capture from manure
decomposition. At voluntary carbon market prices of
8--45~USD/t~CO\textsubscript{2}e~\cite{EcosystemMarketplace2024},
this represents potential annual carbon credit revenue of
\$0.66--3.7~million.

\section{Discussion}

\subsection{Technical Feasibility}

The 37 MW thermal CSP operates at 19.67\% capacity factor, conservative for the site's 4.60 kWh/m²/day DNI. The 70 MWh thermal storage provides approximately 14 hours full-load turbine operation at 5 MW, enabling evening dispatch. Parabolic trough CSP is mature with over 5 GW deployed globally \cite{IRENA2013} and proven in tropical climates like Soroti's \cite{WorldBank2007}.

Conservative biogas production of 30,000 m³/day provides 65\% fuel margin above 3.0 MW average CCGT requirement, accounting for seasonal variation and providing operational headroom. Biogas yield assumptions (0.25 m³ CH4/kg volatile solids) are conservative relative to literature ranges of 0.20-0.40 m³/kg \cite{nzila2012, menardo2012}.

Battery operation at 9 cycles annually (87.6\% of constraint) is well within lithium iron phosphate capabilities exceeding 5,000 cycles at 80\% depth of discharge. The 10 MWh capacity provides 1.4 hours of peak load backup during outages.

\subsection{Sensitivity Analysis}

\begin{figure}
\centering
\includegraphics[width=0.5\textwidth]{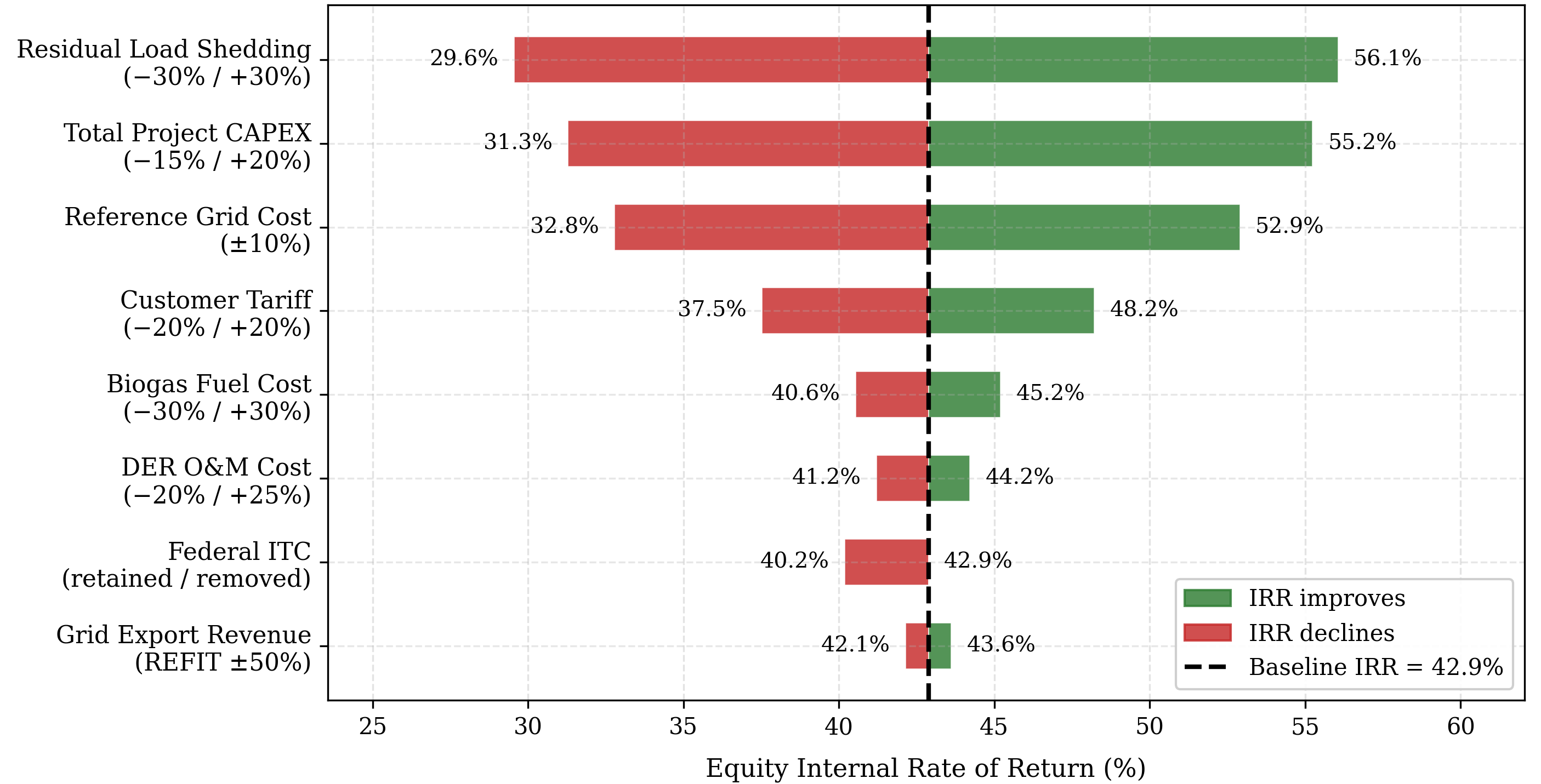}
\caption{One-at-a-time IRR sensitivity to 8
key model parameters. Bar end-points show resulting IRR under
pessimistic and optimistic assumptions for each parameter. Residual load-shedding cost and total project CAPEX are the dominant drivers; IRR remains above 29\% across all individual stress tests.}
\label{fig:tornado}
\vspace{-6pt} 
\end{figure}

The 42.91\% IRR and 2-year payback reflect three compounding drivers
unique to this context: (1) the 69.7\% OPEX savings from on-site
generation displacing \$18.0M/yr in reference grid costs, which include
substantial load-shedding opportunity costs given documented grid
unreliability in Eastern Uganda \cite{ERA2023}; (2) the 30\% investment
tax credit applied in Year~1 (\$1.67M); and (3) export revenue at
REFIT Phase~5 rates. To assess robustness, one-at-a-time sensitivity
analysis was performed across eight parameters (Fig.~\ref{fig:tornado}).

As shown in Fig.~\ref{fig:tornado}, residual load-shedding cost and
total CAPEX have the largest individual influence on IRR, with
customer electricity tariff third. Even under the
most pessimistic combined scenario (20\% CAPEX overrun, 20\% lower
customer tariff, full ITC removal), the IRR remains at 25.2\%, above
the 20\% hurdle rate typical for East African renewable energy
infrastructure~\cite{ERA2023}. Under the optimistic scenarioan (15\%
CAPEX savings, 20\% higher tariff), IRR reaches 61.5\%.

A 10,000-trial Monte Carlo simulation varying all eight parameters
simultaneously was conducted to assess joint uncertainty. Parameters
were sampled from independent triangular distributions centered on
their base-case values with minimum/maximum bounds corresponding to
the pessimistic/optimistic ranges defined in the tornado analysis;
triangular distributions were chosen given the absence of historical
project data for Uganda CSP deployments. The simulation yields a mean
IRR of 42.8\% with a 90th-percentile range of 30.2\%--55.8\%, and
assigns P(IRR~$>$~20\%)~$=$~99.6\% and P(NPV~$>$~0)~$=$~100\%
across all trials.

\subsection{Economic Considerations and Limitations}

Several factors temper the financial results. First, the \$7.9M/yr
non-emergency load-shedding cost (64.9\% of OPEX) reflects 8.3~GWh/yr
of unserved energy (20.4\% of demand); eliminating load shedding would
require additional capacity and substantially reduce the IRR. Second,
the model assumes immediate full customer uptake, whereas actual
ramp-up over 2--3~years would moderate early returns and extend payback
to 3--4~years. Third, currency risk (UGX/USD depreciation), political
risk, and household connection costs (150--250~USD/connection vs.\
rural incomes of $\sim$50~USD/month) are not modeled, and likely
require concessional financing mechanisms. The headline IRR therefore
represents an upper bound; site-specific due diligence is required.

\subsection{Scalability, Replicability, and Limitations}

SIREM requires moderate land area ($\sim$7~hectares), consistent solar
resource (DNI~$>$~4~kWh/m\textsuperscript{2}/day), and organic waste
feedstock, conditions typical in sub-Saharan Africa's agricultural
belt. Biogas collection efficiency of 45\% is conservative and
improvable through dedicated infrastructure at cattle pens and
cooperatives, supported by evidence that larger herds reduce dis-adoption rates~\cite{Lwiza2017}. Successful replication requires IPP regulatory frameworks,
feed-in tariff or PPA mechanisms, project finance below 15\%, and
sufficient peak demand ($\geq$5~MW). Uganda's ERA framework and
comparable East African Community regulations are favourable;
declining CSP costs toward 2030 targets will further improve returns.

Beyond the economic caveats above, the following technical limitations
apply. The 40.8~GWh demand projection assumes steady-state consumption; load growth sensitivity of $\pm$20\% is recommended for pre-feasibility studies. Biogas availability requires pilot digester validation given uncertainties in cattle population dynamics and seasonal collection. Grid interaction modeling does not characterize Soroti substation capacity for 5--7~MW reverse power flow; coordination with UETCL and UEDCL is required to confirm export feasibility.

\section{Conclusion}

This paper demonstrates the technical and economic viability of SIREM,
a hybrid CSP-biogas-battery system to increase electricity access in Eastern Uganda. The optimal configuration (37~MW thermal CSP, 70~MWh TES, 5~MW CSP turbine, 6~MW biogas CCGT, 10~MWh battery) generates
61.3~GWh annually, delivering 32.5~GWh to on-site loads and exporting
28.8~GWh to the grid, with a generation-based LCOE of
0.1991~USD/kWh, 42.91\% IRR, 2-year payback, and 48.8~million USD
NPV over 25~years. This yields 32.2\% cost savings vs. grid-only
supply. Sensitivity analysis across 10,000 Monte Carlo trials confirms
P(IRR~$>$~20\%)~$=$~99.6\%, demonstrating financial robustness under
combined parameter uncertainty. Net system CO\textsubscript{2} is
11,665~metric~tons/year, 2.3\% above the grid-only reference baseline. The primary climate contribution is preventing
82,500~metric~tons CO\textsubscript{2}e/yr in fugitive methane
emissions, supporting Uganda's NDC targets.

SIREM offers a replicable model combining proven technologies with
locally available resources across sub-Saharan Africa's agricultural
regions. Future work should address ownership structure (PPP, private
IPP, or community cooperative), tariff-setting balancing affordability
and cost recovery, reliability constraints to reduce load shedding,
and stakeholder engagement with Soroti District government, customers,
agricultural cooperatives, and civil society.

\section*{Acknowledgment}

The authors thank CMU-Africa for computational resources and Xendee for the microgrid optimization software. Thanks to the NASA LaRC POWER Project and Uganda's Bureau of Statistics, Electricity Regulatory Authority, and Ministry of Energy and Mineral Development for providing data.

\end{document}